\documentclass[iop,twocolumn,tighten]{aastex61}

\usepackage{amssymb}
\usepackage{amsmath, mathrsfs}
\usepackage{amsfonts}
\usepackage{calc}
\usepackage{url}
\definecolor{darkgreen}{rgb}{0,.5,0}

\newfont{\nf}{cmfib8 at 10pt}

\newcommand{\effectiveness}{98.3\%}

\newcommand{\numNotTransitLike}{3274}
\newcommand{\numUncaughtFa}{56}
\newcommand{\numPc}{67}
\newcommand{\reliabilityPercent}{16.0\%}

\newcommand{\nasa} {NASA Ames Research Center, Moffett Field, CA 94035, USA}
\newcommand{\seti}{SETI Institute, 189 Bernardo Ave, Suite 200, Mountain View, CA 94043, USA}
\newcommand{\stsci}{Space Telescope Science Institute, 3700 San Martin Drive, Baltimore, MD 21218}
\newcommand{\bishop}{Dept. of Physics and Astronomy, Bishop's University, 2600 College St., Sherbrooke, QC, J1M 1Z7, Canada}
\newcommand{\mitadd}{MIT Kavli Institute for Astrophysics and Space Research, 77 Massachusetts Avenue, 37-241, Cambridge, MA 02139}

\newcommand{\Rearth}{\,R$_{\oplus}$}

\newcommand{\teff}{\ensuremath{T_{\mathrm{eff}}} }
\newcommand{\logg}{$\log{g}$ }

\newcommand{\kepler}{{\it Kepler}}

\shorttitle{Kepler-452b}
\shortauthors{Mullally et al.}

\begin{document}

\title{Kepler's Earth-like Planets Should Not Be Confirmed Without Independent Detection: The Case of Kepler-452\MakeLowercase{b}}
\author{Fergal Mullally}
\affiliation{\seti}
\altaffiliation{email: fergal.mullally@gmail.com}

\author[0000-0001-7106-4683]{Susan E. Thompson}
\affiliation{\seti}
\affiliation{\nasa}
\affiliation{\stsci}

\author[0000-0003-1634-9672]{Jeffrey L. Coughlin}
\affiliation{\seti}
\affiliation{\nasa}

\author{Christopher J. Burke}
\affiliation{\seti}
\affiliation{\nasa}
\affiliation{\mitadd}

\author[0000-0002-5904-1865]{Jason F. Rowe}
\affiliation{\bishop}

\begin{abstract}
\replaced{We show that the claimed confirmed planet Kepler-452b can not be confirmed using a purely statistical validation approach. \kepler\ detects many more periodic signals from instrumental effects than it does from transits, and it is likely impossible to confidently distinguish the two types of event at low signal-to-noise. In particular, we show that the claimed confirmed habitable-zone planet Kepler-452b (a.k.a. K07016.01, KIC\,8311864) has not been shown to be astrophysical in origin with sufficient confidence, and must still be considered a candidate planet.
A similar analysis applies to other long-period, low SNR, confirmed planets.
}{
We show that the claimed confirmed planet Kepler-452b (a.k.a. K07016.01, KIC\,8311864) can not be confirmed using a purely statistical validation approach. \kepler\ detects many more periodic signals from instrumental effects than it does from transits, and it is likely impossible to confidently distinguish the two types of event at low signal-to-noise. As a result, the scenario that the observed signal is due to an instrumental artifact can't be ruled out with 99\% confidence, and the system must still be considered a candidate planet. We discuss the implications for other confirmed planets in or near the habitable zone.
}

\end{abstract}


\setlength{\parskip}{1.5ex plus0.5ex minus0.2ex}

\section{Introduction}
\kepler's results will cast a long shadow on the field of exoplanets. 
The abundance of exo-Earths derived from \kepler\ data will dictate the design of future direct detection missions.
An oft neglected component  of occurrence rate calculations is an estimate of the reliability of the underlying catalog. As frequently mentioned in \kepler\ catalog papers \citep{Batalha13, Burke14, Rowe15, Mullally15cat, Coughlin16, Thompson17}, not every listed candidate is actually a planet. Assuming they are all planets leads to an over-estimated planet occurrence rate \citep[see, e.g., \S~8 of][]{Burke15}. This problem is especially pronounced for small ($\lesssim 2$\,\Rearth), long-period ($>$ 200\,days) candidates, where the signals due to instrumental effects dominate over astrophysical signals.

External confirmation of \kepler\ planets as an independent measure of reliability is therefore an important part of \kepler's science. While progressively more sophisticated approaches to false positive identification have been employed by successive \kepler\ catalogs, it is not possible to identify every false positive with \kepler\ data alone. Independent, external measures of catalog reliability are a necessary step towards obtaining the most accurate occurrence rates, as well as defining a set of targets for which follow-up observations can be planned. Also, the intangible benefit of being able to point to specific systems that host planets, not just candidates, should not be understated. Kepler-452b is especially interesting in this regard. The target star is similar to the sun, the orbital period is close to one Earth year, and its measured radius ($1.6 \pm 0.2$ \Rearth) admits the possibility of a predominantly rocky composition \citep{Wolfgang15, Rodgers15}.

While small, short-period planets can be confirmed with radial velocities \citep[see, e.g.,][]{Marcy14}, longer period planets are hard to detect in this manner.  
\citet{Torres11}, building on work by \citet{Brown03} and \citet{Mandushev05}, introduced the technique of statistical confirmation, which used all available observational evidence (including high resolution imaging to identify possible background eclipsing binaries) to rule out various false positive scenarios involving eclipsing binaries to the level that the probability the observed signal was a planet exceeds the probability of an eclipsing binary by a factor of 100 or more. \citet{Morton16} used a simplified technique which could be tractably run on the entire candidate catalog, finding and confirming 1284 new systems. \citet{Lissauer12} deduced that the presence of two or more planet candidates around a single star meant that neither was likely to be a false positive and first suggested that a false alarm probability of $<$1\% was sufficient to claim confirmation of a candidate. \citet{Rowe14multis} used Lissauer's argument to claim statistical confirmation of 851 new planets. Rowe et al. and Morton et al. together confirmed the vast majority of confirmed \kepler\ planets.

\replaced{However, in this paper we argue that the statistical confirmation of long-period, low SNR planets is not yet possible. We use Kepler-452b \citep{Jenkins15} as a concrete example, but our argument may also apply to other statistically confirmed  small planets in or near the habitable zone of solar-type stars. The prospect for statistical confirmation of additional Earth-like candidates from \kepler\ is similarly bleak}{However, in this paper we argue that statistical confirmation of long-period, low SNR planets is considerably more challenging than previously believed. 
We use Kepler-452b \citep{Jenkins15} as a concrete example of a planet that should not be considered confirmed to illustrate our argument, but our argument may also apply to other statistically confirmed  small planets (or candidates) in or near the habitable zone of solar-type stars.}. 
Our argument relies on two lines of evidence. First, the number of instrumental false positive (hereafter called false alarm) signals seen in \kepler\ data before vetting dominates over the number of true planet signals at long period and low signal-to-noise (SNR). Second, it is difficult (if not impossible) to adequately filter out enough of those false alarms while preserving the real planets, even with visual inspection. The sample of planet candidates in this region of parameter space is sufficiently diluted by undetected false alarms that it may be impossible to conclude that any single, small, long period planet candidate is not an instrumental feature with a confidence greater than 99\% without independent observational confirmation that the transit is real. \S7.3 of the final \kepler\ planet candidate catalog \citep{Thompson17} discusses the tools and techniques necessary to study catalog reliability, and we draw heavily from that analysis.

\begin{figure*}
     \begin{center}
    \includegraphics[angle=0, scale=.35]{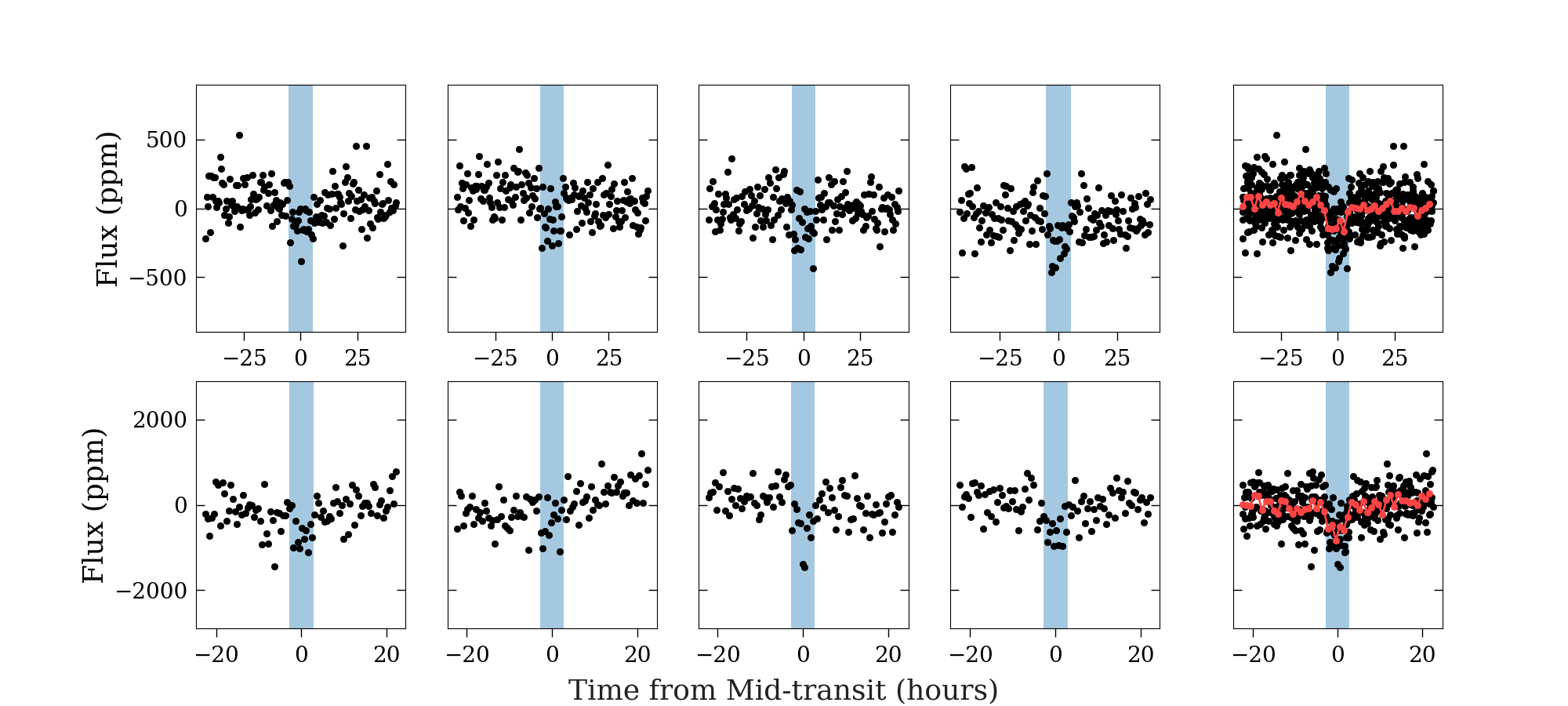}
    \caption{Each panel shows a single transit from a TCE discovered in one of our data sets. The time scale in each panel is hours from mid-transit, with the points the \kepler\ pipeline considers ``in-transit'' highlighted with the vertical blue bar. The panel at far right combines all four events in a folded lightcurve (the red square symbols show the binned folded transit). One row shows transits from a known false alarm TCE found in the inversion run, the other from Kepler-452b. It is not clear from visual examination which one is a claimed planet, and which one is a known systematic signal. (We provide the answer at the end of the paper.)
    \label{invV452}}
     \end{center}
 \end{figure*}

\section{The Limits of Statistical Confirmation}
The basic approach for statistical validation is to compute the odds-ratio between the posterior likelihood that an observed signal is due to a planet, to the probability of some other source.
The Blender approach \citep{Torres11}, used by \citet{Jenkins15} to confirm Kepler-452b, computes

\begin{equation}
\frac{ P(\mathrm{planet})} { P(\mathrm{planet}) + P(\mathrm{EB})}
\end{equation}

\noindent
where $P(\mathrm{EB})$ is the posterior probability the signal is due to an eclipsing binary, and the denominator sums to unity. 
\citet{Morton16} expanded on this approach by fitting models of two types of non-transit signals to their data, effectively computing

\begin{equation}
\frac{ P(\mathrm{planet})} { P(\mathrm{planet}) + P(\mathrm{EB}) + P(\mathrm{FA})}
\end{equation}

\noindent
where FA indicates a false alarm due to a non-astrophysical signal. Notably, they assume the priors on those false alarm models to be low, reflecting their belief that only a small fraction of their sample were non-astronomical signals.

For high SNR detections, or for candidates with periods between 50--200 days, $P(\mathrm{FA})$ is indeed small, and can essentially be ignored. However, \citet{Mullally15cat} first noted the same is not true for candidates with periods longer than $\sim$ 300\,days and clustered close to the SNR threshold for detection (defined as Multiple Event Statistic, MES, $> 7.1$). They urged caution in interpreting those candidates as evidence that there was an abundance of planets between 300 and 500\,days. 

The difficulty in distinguishing low signal-to-noise planets from false alarms is illustrated in Figure~\ref{invV452}. One of the panels shows individual transits from Kepler-452b, and the other from a simulated dataset mimicking the noise properties of \kepler\ data \citep{Coughlin17scramble}. (We summarize how simulated data were created in \S\ref{reliability}.)
Both are detected as TCEs (Threshold Crossing Events, or periodic dips in a lightcurve) by the \kepler\ pipeline \citep{Twicken16}.

It is not obvious, even to the eye, which is the real planet, and which is the artifact. When planets and false alarms look so similar there is no definitive test to distinguish the two. Any vetting process must trade-off between {\it completeness}, the fraction of bona-fide planets correctly identified (or the true positive rate) and {\it effectiveness}, the fraction of false alarms correctly identified (or the true negative rate).

Most of these false alarms are at long period and low SNR. At long periods,
a few systematic ``kinks'' in a lightcurve can often line up, sometimes with low-amplitude stellar variability, to produce a signal that looks plausibly like a transit. These chance alignments are common for 3-4 transit TCEs, but their frequency declines with 5 or more transits. Candidates with 5 or more transits are less likely to be false alarms because the probability of getting 5 events to line up periodically is smaller \citep[see Figure 9 of ][ for more details]{Thompson17}.
The combination of less than ideal effectiveness at low SNR (because transits and false alarms are difficult to tell apart), and an over-abundance of false alarms at long period, leads to a surprisingly low catalog {\it reliability}, or the fraction of claimed planet candidates actually due to transits, and not due to instrumental effects.

\section{The Reliability of Earth-like Planet Candidates }
\label{reliability}
The catalog of \citet{Thompson17} uses a number of tests, collectively called the Robovetter \citep{Coughlin16, Thompson15lpp, Mullally16marshall},  to identify false positive and false alarm TCEs. Each test must be tuned to balance the competing demands of completeness (that as many true planets as possible are passed by each test) and effectiveness (that as many false positives of the targeted type as possible are rejected by the test). The large number of tests applied means that each test must be tuned for high completeness, or the number of bona-fide planets incorrectly rejected quickly becomes unacceptably large.

To test the completeness, effectiveness, and reliability of the planet candidate catalog, the \kepler\ mission produced three synthetic data sets. \citet{Christiansen17ksci}  measured completeness by injecting artificial transits into \kepler\ lightcurves and seeing how many were recovered by the \kepler\ pipeline. 
Producing a sample of transit free lightcurves with realistic noise and systematic properties was more difficult.
The mission tried two approaches, each with their own strengths and weaknesses, as discussed in more detail in \citet{Thompson17}. The first approach (called ``inversion'') was to invert the lightcurves (so that a flux decrement becomes an excess) 
The second approach was to shuffle the lightcurves in time, or ``scramble'' them, to remove the phase coherence of the real transits. All simulated data sets are available for download at the NASA Exoplanet Archive\footnote{Simulated data available at \url{https://exoplanetarchive.ipac.caltech.edu/docs/KeplerSimulated.html}}.

In the discussion that follows, we will consider the performance of the Robovetter for TCEs with periods in the range 200-500\,days and MES \citep[Multiple Event Statistic, or the SNR of detection of the TCE in TPS,][]{Christiansen12} less than 10. This parameter space comfortably brackets Kepler-452b and contains sufficient TCEs to perform a statistical analysis. For brevity, we will refer to this sample as the Earth-like sample because it encompasses the rocky, habitable zone planets \kepler\ was designed to detect \citep{Koch10}. We assume for simplicity that completeness and effectiveness is constant across this box, although in reality the Robovetter performance declines with increasing period and decreasing MES.

 \begin{figure*}[t]
     \begin{center}
    \includegraphics[angle=0, scale=.5]{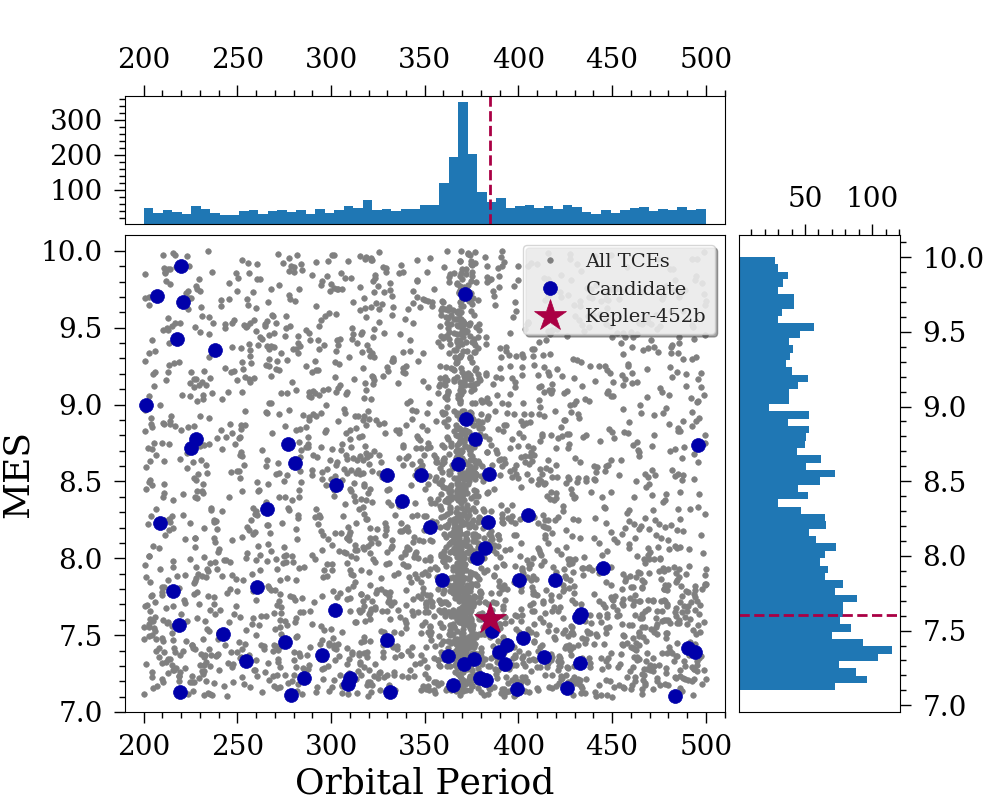}
    \caption{Distribution of planet candidates (large blue circles) and non-candidate TCEs (small grey circles) from the DR25 catalog. The marginal histograms in period (and MES) are shown in the top (and right) panels. The rejected population is dominated by TCEs caused by instrumental systematics in the lightcurves. Because there are so many of these instrumental false alarms, the small number that slip through vetting are a significant fraction of the planet candidate catalog in this region of parameter space. The reliability of any individual planet candidate (or the probability it is due to an astrophysical event and not an instrumental feature) is too low to allow any single candidate to be confirmed without additional observational evidence. \deleted{ The problem is particularly acute near 372\,days, the orbital period of the spacecraft.}  The location of Kepler-452b is indicated by the red star.
    \label{opsTces}}
     \end{center}
 \end{figure*}

\citet{Thompson17} ran the Robovetter on a union of TCEs produced by inversion and the first scrambling run \citep{Coughlin17scramble}\footnote{Three scrambling runs were created, but we only use the first one for consistency with \citet{Thompson17}}. They measure effectiveness as the fraction of these simulated TCEs correctly identified as false alarms in the given parameter space.
The Robovetter performs extremely well on these simulated TCEs, with an effectiveness of \effectiveness. For every thousand false alarms, the Robovetter correctly identifies 983 of them and mis-classifies only 17 as planet candidates.

The original observations (i.e. the real data) produced a huge number of (mostly false alarm) Earth-like TCEs, as shown in Figure~\ref{opsTces}. 
The Robovetter examined 3341 TCEs with periods between 200 and 500\,days and MES $<$ 10 in the original data for DR25. Of these, \numNotTransitLike\ are identified as not-transit-like (i.e. unlikely to be either due to a transit or eclipse; not-transit-like is the label given by the Robovetter to TCEs without a realistic transit shape). The measured effectiveness suggests that an additional \numUncaughtFa\ false alarm TCEs were mistakenly labeled as planet candidates.

There are only \numPc\ planet candidates in this region of parameter space, but if \numUncaughtFa\ are likely uncaught false alarms, then only 11 are bona-fide transit signals. The reliability of the sample (the fraction of planet candidates that are actually due to an astrophysical event such as a transit or eclipse) is then 11 / \numPc\ =  \reliabilityPercent.
This low reliability means that there is only a \reliabilityPercent\ chance that any one planet candidate is actually due to a planet transit, and not a systematic event, far lower than the 99\% threshold for statistical confirmation suggested by \citet{Lissauer12}, or the 99.76\% level confidence claimed by \citet{Jenkins15} for Kepler-452b. The astrophysical false positive rate is clearly only part of the puzzle. For Earth-like candidates, the instrumental false alarm rate must be included in order to statistically confirm planet candidates.

\added{Worse, our ability to estimate reliability is limited by the small numbers of simulations in the parameter space of interest. If we assume the number of uncaught false alarms is drawn from a Poisson distribution, our uncertainty in the number of false alarms is 56$\pm$7.5. The uncertainty in the reliability is then 16$\pm$11\%. Not only do we fail to measure a false positive probability of $<$1\%, our current data prevents us from measuring the false positive probability with a precision of $<1$\%.}

{ Our analysis is slightly over-simplified for clarity. A more detailed calculation would correct the number of measured false alarms for the small number of bona-fide transits mistakenly rejected as false alarms. It should also consider the reasons the Robovetter failed simulated transits, as some rejected TCEs were rejected as variable stars, eclipsing binaries etc. However, none of these improvements can hope to raise the measured reliability to $>99$\%. The error in the computed reliability due to these simplifications is likely small compared to the systematic error in measured effectiveness due to the different abundances of false alarms in the original and simulated data. \added{In particular, assuming constant reliability across the sample slightly over-estimates the reliability of the Kepler-452b, as the effectiveness decreases with decreasing numbers of transits.} 

The fidelity of the simulated false alarm populations is discussed \added{in some detail in \S~4.2 of \citet{Thompson17}. Combining the TCEs from the inversion run and one scrambling run produces a distribution of simulated TCEs as a function of period that matches the observed distribution quite well, but not exactly.  
Because the sources of the observed false alarm TCEs is not precisely understood, we can not rule out the possibility that the simulated datasets are over-producing a certain kind of systematic signal that the robovetter performs well at detecting while under-producing a different kind of signal the robovetter struggles with, thereby overestimating our measured reliability. The converse may also be true. The similarity in the period distributions implies that the simulated populations are well matched to the observed ones, but there is still a small systematic error in our measured reliability that we don't yet know how to measure. 
}

Our argument does not directly apply to the majority of statistically confirmed \kepler\ planets. Transits with high SNR are much less likely to be systematics, and much easier to identify as such if they are. The two largest contributions to the set of statistical confirmations, \citet{Rowe14multis} and \citet{Morton16}, both restrict their analysis to candidates detected with a SNR $>10$. { Kepler-452b has the lowest SNR of all the long-period planets with claimed confirmations, and thus has the most tenuous claim to confirmation. However, 99\% reliability is a high threshold to reach at low SNR, and the confidence with which many of the other small, habitable-zone planets were claimed to be confirmed is likely over-stated. We leave the analysis of these systems as future work.}

\subsection{Raising Reliability}
In this section we discuss some refinements to our analysis that increase the estimated reliability of some of the candidates in the Earth-like sample. However none of these refinements increase the reliability to the desired 99\% level. 

\subsubsection{Improved Parameter Space Selection}

Following \citet{Thompson17}, if we restrict our analysis to TCEs around main-sequence stars (4000\,K $< \teff <$ 7000\,K and \logg\ $>$ 4.0) we are looking at signals observed around photometrically quieter stars. The Robovetter performs much better for these stars, and the measured reliability increases to $\approx$ 50\%. \added{ Further restricting the sample to exclude the period range with the highest false alarm rate (360 to 380\,days, as shown in Figure~\ref{opsTces}) increases the reliability to only 54\%.}

This is close to the ceiling for reliability that can be achieved by analyzing smaller parameter spaces. 
Choosing narrower slices of parameter space can produce small gains in effectiveness, but comparatively larger changes in the numbers of candidates. The measured reliability increases slowly (if at all) and becomes noisier, as the effects of small number statistics begin to dominate.

\subsubsection{High score TCEs}
\citet{Thompson17} include a score metric, which attempts to quantify the confidence the Robovetter places in its classification of a TCE as a planet candidate or a false alarm. The score is not the probability that a TCE is a planet.
 High scores ($>0.8$) indicate strong confidence that a TCE is a candidate, while low scores ($<0.2$) indicate high confidence that a TCE is either a false positive or false alarm. An intermediate score indicates that the TCE was close to the threshold for failure in one or more of the Robovetter metrics. Kepler-452b has a score of 0.77. 
 
If we repeat our analysis on the small, long-period TCEs around FGK dwarves, but require a score $\geqslant 0.77$ to consider a TCE a candidate, we find an effectiveness in simulated data of 99.97\%, and a reliability of the catalog (between periods of 200-500\,days and MES $<$ 10) of 92\% based on 9 planet candidates. While this reliability is much stronger than before, it is still some distance from the 99\% reliability required to claim statistical confirmation. The effectiveness estimate is also based on a small number of candidates and simulated TCEs passing the Robovetter. If the number of incorrectly identified simulated TCEs changes by just one, the measured reliability changes by 8\%. Even if the measured reliability for high score TCEs was $>99$\%, the uncertainty in the measured reliability casts sufficient doubt to counter any claimed confirmation.

\subsubsection{Multiple planet systems}
\citet{Lissauer12} first noted that the presence of multiple planet candidates around a single target star was strong evidence that all candidates in that system were planets, and not eclipsing binaries. The argument is essentially that while planet detections and eclipsing binary detections are both rare, multiple detectable planets in a single system are much more likely than an accidental geometric alignment between a planet host system and a background eclipsing binary.  As recognized by \citet{Rowe14multis}, a similar argument does not automatically apply to false positives. Lissauer assumed that planetary systems and eclipsing binaries are distributed uniformly across all targets. Systematic signals come from a variety of sources, some of which are concentrated in specific regions of the focal plane, or types of stars. Some targets are very much more likely to see multiple false alarm TCEs than a uniform distribution would predict.  It may be possible to apply the Lissauer analysis once a richer model of the false alarm distribution is known. 

\subsubsection{Relying on earlier catalogs}
The reliability of a candidate may be higher if it appears in multiple \kepler\ catalogs. The TPS algorithm underwent extensive development between the Q1-Q16 catalog of \citet{Mullally15cat}, the DR24 catalog of \citet{Coughlin16} and the DR25 catalog (the three catalogs to include 4 years of \kepler\ data). The different catalogs could, in principle, be used as quasi-independent detections at the low SNR limit. Unfortunately, due to the lack of diagnostic runs on the earlier catalogs (reliability is unmeasured for DR24, and neither reliability nor completeness for Q1-Q16), it is impossible to quantify the improved reliability of a candidate in this manner. 
For the specific case of Kepler-452b, it was detected in both the DR24 and DR25 catalogs, but was not detected in Q1-Q16, even though all 4 transits were observed. 

\deleted{
We conclude that no planet in our Earth-like sample should be considered a confirmed planet unless it is either confirmed by an independent {\it observational} technique, or a new method is devised to improve the effectiveness of the vetting. Given the difficulties of creating simulated data that reproduces the time-varying, non-Gaussian noise properties of \kepler\ data, and the considerable development effort already invested in the Robovetter, we are doubtful such a method will be devised in the near future. 
}
\explain{Moved to discussion, and conclusions softened}

\section{Discussion \& Conclusion}

\added{We show statistical validation is insufficient to confirm Kepler-452b at the 99\% level, and that Kepler-452b should no longer be considered a confirmed planet. Although use of this threshold to claim confirmation is somewhat arbitrary, it has seen widespread adoption in the community. To our knowledge, there are no claims in the literature of the claimed confirmation of a transiting planet with confidence less than this threshold.}

\added{Our simplest calculation argues that there is only a 16\% chance that the detected signal on the target star is due to a transit, while our most optimistic calculation sets the probability at 92\%. Even this most optimistic assessment is still an order of magnitude less confident than our threshold, and nearly 2 orders less than that claimed in the discovery paper \citep[99.76\%,][]{Jenkins15}. While the exact choice of threshold to claim confirmation may be a matter of taste and convention, setting the threshold low enough to admit Kepler-452b would result in significant contamination of the confirmed planet catalog with false alarms, and negate the goals of creating a confirmed planet list.}

\added{It may be possible to devise a new statistical approach that validates Kepler-452b. However, given the difficulties of creating simulated data that reproduces the time-varying, non-Gaussian noise properties of \kepler\ data, and the considerable development effort already invested in the Robovetter, we are doubtful such a method will be devised in the near future. Instead, we advocate that the transit events should be directly confirmed by independent observations, such as the Hubble follow-up of Kepler-62 by \citet{Burke17hstprop}.}

\added{ The un-confirmation of Kepler-452b is interesting in its own right, given the similarity between its measured parameters and those of the Earth, but our result has implications for other long-period confirmed \kepler\ planets. \citet{Torres17} claimed to statistically confirm 6 habitable-zone candidates originally detected at long-period and MES $<$ 12. These planets were all detected at higher SNR than Kepler-452b, and are therefore less likely to be caused by an instrumental artifact. However, the lack of precision for our reliability estimates in this work does raise a concern for these other systems. The burden of proof for claiming that a planet exists lies in demonstrating that the probability of all other scenarios is definitely less than the chosen threshold. 
 If the likelihood of a transit signal is drawn from a distribution (i.e a posterior) that overlaps significantly with the threshold for acceptance, then that burden of proof is not met. For Kepler-452b, the uncertainty in our measurement of reliability is an order of magnitude higher that what is required for confirmation.}

The inability to confirm individual planet candidates by statistical techniques should not be considered a failure of the \kepler\ mission (or of the statistical techniques themselves, which are applicable when their underlying assumptions remain valid). \kepler's goal was to establish the frequency of Earth-like planets in the Galaxy, and the conclusion that at least 2\% 
(and possibly as many as 25\%) of stars host an Earth-size planet in the shorter-period range of 50--300\,days \citep{Burke15}  is not strongly affected by this result. The reliability of the \kepler\ sample in this regime is much higher, and sensitivity of occurrence rate calculations is less sensitive to catalog reliability. It will be necessary to account for catalog reliability to properly extend occurrence rate calculations out toward the longer periods that encompass the habitable zone around G type stars.

\acknowledgements
\added{We thank the referee, Tim Brown, for his constructive comments.}
This work is supported by NASA under grant number NNX16AJ19G.
Some of the data presented in this paper were obtained from the Mikulski Archive for Space Telescopes (MAST). STScI is operated by the Association of Universities for Research in Astronomy, Inc., under NASA contract NAS5-26555. Support for MAST for non-HST data is provided by the NASA Office of Space Science via grant NNX09AF08G and by other grants and contracts.
This research has also made use of the NASA Exoplanet Archive, which is operated by the California Institute of Technology, under contract with the National Aeronautics and Space Administration under the Exoplanet Exploration Program.
Figure~\ref{invV452} shows Kepler-452b in the top panel, and the inverted TCE 11961208-01 (period 425\,days, best fit radius of 2\,\Rearth) in the lower panel.

{\it Facilities}: \kepler


\begin{thebibliography}{}
\expandafter\ifx\csname natexlab\endcsname\relax\def\natexlab#1{#1}\fi
\setlength{\itemsep}{0ex}
\small
\bibitem[{Batalha et~al.(2013)}]{Batalha13}Batalha, N.~M., Rowe, J.~F., Bryson, S.~T., et~al. 2013, \apjs, 204, 24

\bibitem[{Brown(2003)}]{Brown03}Brown, T.~M. 2003, \apjl, 593, L125

\bibitem[{Burke(2017)}]{Burke17hstprop}Burke, C. 2017, ``Completing Kepler's Mission to Determine the Frequency of Earth-like Planets'', HST Proposal 15129

\bibitem[{Burke et~al.(2014)}]{Burke14}Burke, C.~J., Bryson, S.~T., Mullally, F., et~al. 2014, \apjs, 210, 19

\bibitem[{Burke et~al.(2015)}]{Burke15}Burke, C.~J., Christiansen, J.~L., Mullally, F., et~al. 2015, \apj, 809, 8

\bibitem[{Christiansen(2017)}]{Christiansen17ksci}Christiansen, J.~L. 2017, {\it {Kepler Science Document, KSCI-19110-001}}

\bibitem[{Christiansen et~al.(2012)}]{Christiansen12}Christiansen, J.~L., Jenkins, J.~M., Caldwell, D.~A., et~al. 2012, \pasp, 124, 1279

\bibitem[{Coughlin(2017)}]{Coughlin17scramble}Coughlin, J.~L. 2017, {\it {Kepler Science Document, KSCI-19114-002}}

\bibitem[{Coughlin et~al.(2016)}]{Coughlin16}Coughlin, J.~L., Mullally, F., Thompson, S.~E., et~al. 2016, \apjs, 224, 12

\bibitem[{Jenkins et~al.(2015)}]{Jenkins15}Jenkins, J.~M., Twicken, J.~D., Batalha, N.~M., et~al. 2015, \aj, 150, 56

\bibitem[{Koch et~al.(2010)}]{Koch10}Koch, D.~G., Borucki, W.~J., Basri, G., et~al. 2010, \apjl, 713, L79

\bibitem[{Lissauer et~al.(2012)}]{Lissauer12}Lissauer, J.~J., Marcy, G.~W., Rowe, J.~F., et~al. 2012, \apj, 750, 112

\bibitem[{Mandushev et~al.(2005)}]{Mandushev05}Mandushev, G., Torres, G., Latham, D.~W., et~al. 2005, \apj, 621, 1061

\bibitem[{Marcy et~al.(2014)}]{Marcy14}Marcy, G.~W., Isaacson, H., Howard, A.~W., et~al. 2014, \apjs, 210, 20

\bibitem[{Morton et~al.(2016)}]{Morton16}Morton, T.~D., Bryson, S.~T., Coughlin, J.~L., et~al. 2016, \apj, 822, 86

\bibitem[{Mullally et~al.(2015)}]{Mullally15cat}Mullally, F., Coughlin, J.~L., Thompson, S.~E., et~al. 2015, \apjs, 217, 31

\bibitem[{Mullally et~al.(2016)}]{Mullally16marshall}------. 2016, \pasp, 128, 074502

\bibitem[{Rogers(2015)}]{Rodgers15}Rogers, L.~A. 2015, \apj, 801, 41

\bibitem[{Rowe et~al.(2014)}]{Rowe14multis}Rowe, J.~F., Bryson, S.~T., Marcy, G.~W., et~al. 2014, \apj, 784, 45

\bibitem[{Rowe et~al.(2015)}]{Rowe15}Rowe, J.~F., Coughlin, J.~L., Antoci, V., et~al. 2015, \apjs, 217, 16

\bibitem[{Thompson et~al.(2017)}]{Thompson17}Thompson, S.~E., Coughlin, J.~L., Hoffman, K., et~al. 2017, ArXiv:1710.06758

\bibitem[{Thompson et~al.(2015)}]{Thompson15lpp}Thompson, S.~E., Mullally, F., Coughlin, J., et~al. 2015, \apj, 812, 46

\bibitem[{Torres et~al.(2011)}]{Torres11}Torres, G., Fressin, F., Batalha, N.~M., et~al. 2011, \apj, 727, 24

\bibitem[{Torres et~al.(2017)}]{Torres17}Torres, G., Kane, S.~R., Rowe, J.~F., et~al. 2017, \aj, 154, 264

\bibitem[{Twicken et~al.(2016)}]{Twicken16}Twicken, J.~D., Jenkins, J.~M., Seader, S.~E., et~al. 2016, \aj, 152, 158

\bibitem[{Wolfgang et~al.(2015)}]{Wolfgang15}Wolfgang, A., \&  Lopez, E. 2015, \apj, 806, 183

\end{thebibliography}


\end{document}